\newif\ifpublish
\publishfalse

\documentclass[conference]{IEEEtran}
\IEEEoverridecommandlockouts
\usepackage{cite}
\usepackage{amsmath,amssymb,amsfonts,amssymb}
\usepackage[mathscr]{euscript}
\newmuskip\pFqmuskip

\usepackage{algorithmic}
\usepackage{graphicx}
\usepackage{textcomp}
\usepackage{xcolor}
\usepackage{xspace}
\usepackage[inline]{enumitem}

\usepackage{dblfloatfix}
\usepackage[caption=false,font=normalsize,labelfont=sf,textfont=sf]{subfig}
\usepackage{pgfplots}
\usepackage{tikz}
\usetikzlibrary{chains,decorations.pathreplacing}
\tikzstyle{arrow} = [thick,->,>=stealth]
\def\BibTeX{{\rm B\kern-.05em{\sc i\kern-.025em b}\kern-.08em
    T\kern-.1667em\lower.7ex\hbox{E}\kern-.125emX}}
\usetikzlibrary{positioning, shapes, arrows}

\newcommand*\pFq[6][8]{%
	\begingroup % only local assignments
	\pFqmuskip=#1mu\relax
	\mathcode`\,=\string"8000
	\begingroup\lccode`\~=`\,
	\lowercase{\endgroup\let~}\pFqcomma
	{}_{#2}F_{#3}{\left[\genfrac..{0pt}{}{#4}{#5};#6\right]}%
	\endgroup
}
\newcommand{\pFqcomma}{\mskip\pFqmuskip}
\providecommand{\name}{\textnormal{Yggdrasil}\xspace} %name of our alogrithm, use macro to avoid misspelling 

\providecommand{\setup}{\ensuremath{{\sf{SetUp}}}\xspace} %system setup algorithm
\providecommand{\user}{\ensuremath{{\sf Client}}\xspace} %client-side algorithm
\providecommand{\cloud}{\ensuremath{{\sf Cloud}}\xspace} %server-side algorithm

\providecommand{\deletions}{\ensuremath{{\sf{Del}}}\xspace} %client DELETIONS

\providecommand{\swap}{\ensuremath{{\sf{Swap}}}\xspace} %cloud SWAPPING

\providecommand{\compress}{\ensuremath{{\sf{Comp}}}\xspace} 
\providecommand{\decompress}{\ensuremath{{\sf{Deco}}}\xspace} 

\providecommand{\upload}{\ensuremath{{\sf{Upload}}}\xspace} %user upload
\providecommand{\get}{\ensuremath{{\sf{Get}}}\xspace} %user retrieval
\providecommand{\baseF}{\ensuremath{r}\xspace}  % Fraction of number of basis
\providecommand{\nodel}{\ensuremath{x}\xspace}  % number of deletions
\providecommand{\sizefid}{\ensuremath{s_{\id}}\xspace}  % 
\providecommand{\nof}{\ensuremath{f}\xspace}  % number of recodrs
\providecommand{\nob}{\ensuremath{b}\xspace}  % number of bases
\providecommand{\system}{\ensuremath{\mathscr{S}}\xspace} %cloud generalized deduplication 
\providecommand{\base}{\ensuremath{{\sf{base}}}\xspace} %base
\providecommand{\baseset}{\ensuremath{\mathcal{S}}\xspace} %set of all basis
\providecommand{\id}{\ensuremath{{\sf fid}}\xspace} %number of basis

\providecommand{\deviation}{\ensuremath{ {\sf dev}}\xspace} %deviation
\providecommand{\dist}{\ensuremath{ {\sf dist}}\xspace} %deviation
\providecommand{\threshold}{\ensuremath{\tau}\xspace} %deviation

\providecommand{\hit}[1]{\ensuremath{\overline{#1}}} %file / chunck / string

\providecommand{\file}{\ensuremath{F}\xspace} %file / chunck / string
\providecommand{\policy}{\ensuremath{{\sf Policy}}\xspace} %policy that describes the distribution of the set of bases

\providecommand{\local}{\ensuremath{D}\xspace} %data locally stored by clinets

\providecommand{\outsource}{\ensuremath{\file'}\xspace} %data outsorced to the cloud

\providecommand{\database}{\ensuremath{DB}\xspace} %outsourced database

\providecommand{\compratio}{\ensuremath{\mathcal{C}}\xspace} %outsourced database
\providecommand{\ucr}{\ensuremath{\compratio_\user}\xspace} 
\providecommand{\ccr}{\ensuremath{\compratio_\cloud}\xspace} 
\providecommand{\unmetric}{\ensuremath{\mathcal{U}}\xspace} 
\providecommand{\distribution}{\ensuremath{\mathcal{D}}\xspace}

\providecommand{\size}{\ensuremath{{{\sf size}}}} %outsourced database
\providecommand{\adversary}{\ensuremath{\mathcal{A}}\xspace} 

\providecommand{\nopreim}{\ensuremath{m }\xspace} 

\providecommand{\changev}{\ensuremath{{\sf ChngV}}\xspace}  
\usepackage[colorinlistoftodos
,disable
]{todonotes}

\begin{document}
\title{Yggdrasil: 
	%Connecting Ground and Cloud Deduplication for Compression and Security in Multi User Setting 
	Privacy-aware Dual Deduplication in Multi Client Settings \vspace{-0.2cm}
}

\author{\IEEEauthorblockN{Hadi Sehat$^{1}$, Elena Pagnin$^{2}$, Daniel E. Lucani$^{1}$}%, Claudio Orlandi$^{3}$}
  \IEEEauthorblockA{$^{1}$Agile Cloud Lab, Department of Engineering, DIGIT, Aarhus University, Aarhus, Denmark\\
    $^{2}$ Department of Engineering and Information Theory, Lund University, Lund, Sweden\\
%    $^{3}$ Department of Computer Science, DIGIT, Aarhus University, Aarhus, Denmark\\
\{hadi,daniel.lucani\}@eng.au.dk,elena.pagnin@eit.lth.se}%, orlandi@cs.au.dk}   
\vspace{-1.5em}
}

%\and
%\and
%\and
%\IEEEauthorblockN{4\textsuperscript{th} Elena Pagnin}
%\IEEEauthorblockA{\textit{dept. Engineering and Information Theory } \\
%\textit{Lund University}\\
%Lund, Sweden \\
%elena.pagnin@eit.lth.se}
%}

\maketitle
\begin{abstract}
  
  This paper proposes \name, a protocol for privacy-aware %, configurable, and distributed 
  dual data deduplication in multi client settings.
%  To achieve this, we exploit the concept of generalized data deduplication, \name considers a family of transformations that splits data chunks into a \emph{basis} and a \emph{deviation} prior to deduplicating the basis. 
%  This transformation has been shown to improve the potential of compression.
%  Our goal is to expand this concept in three significant ways.
\name is designed to reduce the cloud storage space while safeguarding the privacy of the client's outsourced data.
\name combines three innovative tools to achieve this goal.
%Concretely, our solution builds on the footprint of generalized deduplication and brings three significant improvements.
First, generalized deduplication, an emerging technique to reduce data footprint. 
Second, non-deterministic %(e.g., data-dependent, random)  
transformations %. We consider transformations 
that are described compactly and improve the degree of data compression in the Cloud (across users). %, thus reducing the space needed to store outsourced data.
%  This constrast with current work on generalized deduplication, which has focused on deterministic transformations.
Third, data preprocessing %. In \name 
in the clients in the form of lightweight, privacy-driven transformations prior to upload.
This guarantees that an honest-but-curious Cloud service trying to retrieve the client's actual data will face a high degree of uncertainty as to what the original data is. 
%  The uncertainty level is proportional to the fraction of the original data locally stored by the client. 
%  clients carry out a first round of generalized deduplication in a chunk to generate the basis uploaded to the Cloud and only storing a description (deviation) of those transformations locally.
%  Second, we allow the Cloud to perform a second round of (generalized) deduplication to reduce the costs of storage in the Cloud itself across multiple clients.
%  Third, we consider non-deterministic (e.g., data-dependent, random) transformations that can be described compactly. This constrast with current work on generalized deduplication, which has focused on deterministic transformations.
  We provide a mathematical analysis of the measure of uncertainty as well as the compression potential of our protocol. 
%  Finally, we test \name on a real world data set (HDFS logs). 
  Our experiments with a HDFS log data set shows that $49$~\% overall compression can be achieved, with clients storing only $12$~\% for privacy and the Cloud storing the rest.
  This is achieved while ensuring that each fragment uploaded to the Cloud would have ~$10^{293}$ possible original strings from the client. Higher uncertainty is possible, with some reduction of compression potential. 
  
\end{abstract}


% To allow compilation of the file
% !TeX root = ./../main.tex

\section{Introduction}
\label{intro}
The massive migration of data from  storage facilities on `the Premise' to `the Cloud' has led to a boom in the offer of Cloud Storage Providers (CSPs). 
Nowadays, people can turn to a vast spectrum of CSPs that offer a virtually unlimited storage accessible from `anywhere' in the world. 
To deliver the promised features to multiple clients at a competitive price, CSPs resort to compression techniques that reduce the footprint of data, so as to fit more in less space. 
A popular approach to reach this goal is to adopt \textit{data deduplication} techniques. 
In brief, these solutions work by recognizing whether a freshly uploaded file is \emph{the same} as one already stored on the server.  
If so, there is no need to store the new file: a pointer to the already existing copy would suffice. 
Obviously the chance that clients upload the exact same data is quite low, but applying this technique on chunks of files leads to a non trivial compression capability. 
This relevant 
%There is a relevant impact on storage size 
when a client stores different versions of the same file (thus, files have several common chunks), or different clients use the same kind of files, e.g., virtual machine disk images of different Linux distributions~\cite{Jin2009,Ng2011}

Vestergaard et al. recently proposed \emph{generalized deduplication} (GD)~\cite{Vestergaard2019a,Vestergaard2019b}, a technique to further reduce the footprint of the storage systems. 
The intuition behind GD is to perform data compression on chunks that are \emph{nearly} identical rather than \emph{exactly} the same. 
%In this case, given a file chunk, the cloud server may efficiently identify existing chunks that are close to it, instead of having to explicitly compare to stored chunks. 
This is possible by using a transformation function that maps each chunk to a \emph{basis} and a \emph{deviation}, where different chunks may have the same basis but different and unique deviations to differentiate them.
%Deduplication is carried out on the basis alone.
The system assigns a fingerprint, e.g., a hash function to each basis %and uses this fingerprint to identify whether or not there was a similar chunk in the system already.
and carries out deduplication on the bases.
%If there is, only the deviation to it needs to be stored.
Using GD, similar chunks, i.e., chunks that are mapped to the same basis are deduplicated together.
The system then stores a pointer to the basis and the deviation in form of a small textual deviation with the information on how the new chunk \textit{differs} from the pointed basis. 
It has been shown that generalized deduplication achieves a higher level of compression than classic deduplication techniques~\cite{Vestergaard2019b}.
%\elena{even in privacy-aware multi clients settings ! should we cite our encryption for generalized deduplication paper?}
%\daniel{I think we would need to upload it to arxiv to do this properly - but I am fine with this, we may just be cutting it a bit close}
%\elena{let's keep it in mind in case one of the two papers gets accepted. For the submission we won't cite unpublished work}

Trivially, deduplication techniques perform better on highly correlated data. 
However, similarities appear only if the files are uploaded in plaintext\footnote{Alternatively, the files could be encrytped using a deterministic encryption scheme, and then uploaded. However deterministic encryption cannot be semantically secure.
Moreover to guarantee meaningful deduplication of ciphertexts generated by different clients there needs to be some coordination on the encryption key.}, which is not desirable in many applications. 
Privacy-conscious clients may upload encrypted data to the Cloud. 
The semantic security of the encryption implies that ciphertexts look random. In particular, files that could be deduplicated become uncorrelated when encrypted, undermining the whole purpose of deduplication.

Up to now, the bulk of work performs deduplication for storage on plaintext data at one location, usually on the Cloud side, since it is assumed to have more computational power a client. 
In contrast, we consider the following \emph{unorthodox} setting: secure deduplication is carried out, in a subsequent manner, by two parties. % (in two separate world). 
That is, the client (e.g., user end device, local storage system, private Cloud) and the server (e.g., CSP).
We call this method \emph{dual deduplication}.
We present a solution that allows clients to outsource their data in a privacy-preserving manner, while the server is guaranteed a high compression rate.
%Both client(s) and server may decide disjointly on how to carry out the compression process.

We achieve this solution by letting the clients preprocess their data prior to upload.
% in a GD fashion. 
The outcome of this process is a pair where the first item is an outsourced generalized deduplication friendly `\textit{basis}' that the client sends to the cloud. In practice, this could be achieved by the client uploading a single file containing a number of unique bases of the same size or by uploading each basis separately.
% These basis are required for recovering the original file.
% Note that each basis could be used multiple times by the client due to its internal GD process.
%The size of the bases could be a parameter specified by the Cloud.
The second item is a deviation, a short string that simultaneously serves two purposes: 
(1) enabling the correct recovery of the original data from the outsourced file (e.g., indicating which basis was used and how was it modified to generate the $i$-th data chunk); 
and (2) providing some level of privacy on the outsourced data. 

Having received a set of deduplication-friendly bases, the CSP can use GD to successfully deduplicate bases received from numerous clients, reducing the total storage space to a fraction of what could be achieved from raw, unprocessed data. 
Although the overall storage space needed by the server and the clients may be higher than if deduplication was carried out only on the server side,  
the storage space required by each party, i.e., client and server separately, is considerably smaller compared to plain storage on the client side or generalized deduplication on unprocessed data on the server side.
%we are interested in scenarios where the clients apply some transformation on their data that simultaneously makes them more deduplication friendly on the cloud side and prevents the cloud from learning the data. 
%Receiving such preprocessed data, the cloud storage provider can perform high ratio data compression via the powerful generalized deduplication technique across data uploded by different clients.

We name our solution \name, as the Cosmic Tree of Life in the Norse mythology. 
\name is an enormous ash tree that connects the different worlds with the heavens. 
We use this as a metaphor for our system 
%the different worlds represent multiple clients and the heavens is the Cloud server
(see Fig. \ref{fig:system}). 
%\name provides a way to connect these worlds, its sap is data that flows between the clients and the CSP. 
The clients preprocess the data and keep a fraction of it for privacy reasons. 
The Cloud collects the deduplication-friently bases output by the clients and organizes them into a (compressed) foliage. 

The contributions of this paper are organized as follows. 
Section \ref{sec:model} introduces %the model and performance metrics of \name. 
%Here we describe 
the framework of \name; 
the kind of adversary we deal with (an honest-but-curious CSP) and its goal (to reconstruct clients' original data); 
and the mathematical expressions we use to measure the performance of the system including compression ratios, and the uncertainty metric measuring the privacy retained by clients against our adversary. 
%privacy is  measured as the degree of uncertainty faced by an honest-but-curious CSP while trying to guess the original chunks of an uploaded file that is pre-processed by the client. 
Section \ref{sec:contribution} explains our solution, \name, in detail (its algorithms and how they interact). 
Section \ref{sec:analysis} presents our 
%mathematical analysis of \name. The main contribution here are 
upper bounds on the different compression ratios and privacy analysis of our solution. 
Section \ref{sec:experiments} collects and discusses the numerical results obtained when testing \name on  a real dataset of HDFS log files. 
Our main focus is the compression potential of our proposal in different settings depending on the way we parse raw data and the type of transformations allowed on the clients' side. 
Section \ref{sec:conclusion} concludes the paper and highlights directions for future work.

% To allow compilation of the file
% !TeX root = ./../main.tex
\section{System Model and Performance Metrics}\label{sec:model}

In this section, we define our system model, the attacker model and then present the metrics we use to analyze the performance and privacy of our proposal \name.
%First, we define our system model;  
%second our attacker model; 
%finally we present metrics to analyze the system performance and security. 
%The metrics will be used to evaluate our proposal \name. 

\subsection{System Model}\label{sec:model_system}
\begin{figure}[!t]
	\centering
	\vspace{-0.1em}
\begin{tikzpicture}[font=\small\sffamily, thick, scale=0.61]

\node[cloud, cloud puffs=19, cloud ignores aspect, minimum width=4cm, minimum height=1.4cm, align=center, draw] at (0cm, 0cm) {Cloud};
\node [alias=cloud] at (0,-1.2){};
\draw (-2.2,-4.5) node[anchor=north east,rounded rectangle,
draw, alias=one]{Client$_1$}; 
\path[<->,very thick] (one) edge [bend right] ([xshift=-6pt] cloud);
\draw (0,-4) node[anchor=north,rounded rectangle,
draw, alias=two]{Client$_2$}; 
\path[<->,very thick] (two) edge ([xshift=0pt] cloud);

\draw (1.7,-4.6) node[anchor=north west, rounded rectangle,
draw, alias=three]{Client$_3$}; 
\path[<->,very thick] (three) edge [bend left] ([xshift=6pt] cloud);

\draw (2,-3.2) node[anchor=north west,rounded rectangle,
draw, alias=fuor]{Client$_n$}; 
\path[<->,very thick] (fuor) edge [bend left] ([xshift=12pt] cloud);

\end{tikzpicture}
\vspace{-0.7em}
\caption{\name system model: several independent clients upload data to the same cloud storage provider.}
\label{fig:system}
\vspace{-1.5em}
\end{figure}
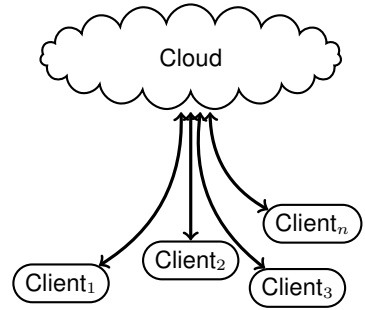

Figure~\ref{fig:system} depicts our system model. 
Clients' desiderata is 
to retain some level of privacy on their files
while minimizing the amount of local storage.
The CSP desiderata is to optimize its storage space. 
In order to meet all desiderata simultaneously, 
we let 
clients apply some transformations on their data, prior to upload. 
Such transformations aim to prevent the CSP (or any third party) from easily guessing the clients' original (raw) data while requiring required minimal storage on the clients' side. 
To minimize the storage requirements on the Cloud side, 
we let the CSP perform Generalized Deduplication. 
To further decrease storage, our model envisions a CSP that processes outsourced data before running GD.

We consider that the system operates on data strings with $k$-bit symbols, i.e., any symbol can take $N = 2^k$ possible values.
A file is broken up into a number of original strings of size $n_o$ symbols, i.e., $\file \in (\{0,1\}^k)^{n_o}$.
After a client applies its transformation, the resulting base (called \outsource) of size $n_b$ symbols
and there is an associated local deviation \local that captures the changes performed on the original string. 
%All the original strings $F$ form \database.
%Prior to upload, \user transforms \file into a shorter and fix-length string $\outsource\in (\{0,1\}^k)^{n_b}$ for  \cloud; and a local deviation \local that is stored on the client machine. 
At a given point, there are $\nof$ strings in the \user and $\nob$ bases stored in the \cloud.

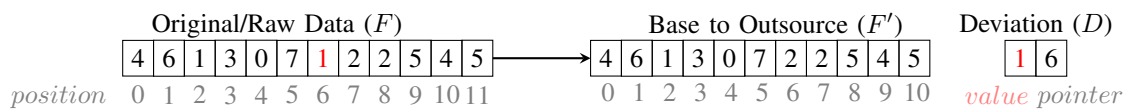
\begin{figure*}[t]
	\centering
	\begin{tikzpicture}[
	node distance=0pt,
	start chain = A, %going right,
	X/.style = {rectangle, draw,% styles of nodes in string (chain)
		minimum width=2ex, minimum height=3ex,
		outer sep=0pt, on chain},
	B/.style = {decorate,
		decoration={brace, amplitude=5pt,
			pre=moveto,pre length=1pt,post=moveto,post length=1pt,
			raise=1mm,
			#1}, % for mirroring of brace, if necessary
		thick},
	B/.default=mirror, % by default braces are mirrored
	]
	\foreach \i in {4,6,1,3,0,7}
	\node[X,right] {\i};

		\node[X,text=red,right] {1};
	
	\foreach \i in {2,2,
		5,4}
	\node[X,right] {\i};
	
	\node (end)[X, right] {5};
	\node (begin)[X, right = 1.3cm] {4};
	
	\foreach \i in {6,1,3,0,7,2,2,
		5,4,5}
	\node[X, right] {\i};
	
	\node[X,,text=red, right = 1 cm]{1};
	
	\node[X, right] {6};
	
	\draw [arrow] (end)  -- (begin);
	\node (B1) [inner sep=1pt,above=of A-6.north west] {Original/Raw Data ($\file$)};
	\node (B2) [inner sep=1pt,above=of A-19.north west] {Base to Outsource ($\file'$)};
	\node (B3) [inner sep=1pt,above=of A-25.north west] {Deviation ($D$)};
		\foreach \x [count=\i] in {0, ...,11}
		\node [below=of A-\i]{\textcolor{black!50}{\x}};
		\node[below=of A-1, xshift=-30pt]{\textcolor{black!50}{$position$}};
		\foreach \x  in {0, ...,10}
		\pgfmathsetmacro{\i}{\x+13}
		\node [below=of A-\i,xshift=-5pt, yshift=-6pt]{\textcolor{black!50}{\x}};
		\node [below=of A-24,xshift=-9pt]{\textcolor{red!50}{{$value$}}};
		\node [below=of A-25,xshift=11pt ]{\textcolor{black!50}{$pointer$}};

		\end{tikzpicture}
		\vspace{-1em}
	\caption{Example of a 1-deletion: the random index is $i=6$ and the corresponding deleted value is 1 (highlighted in red).}
	\label{dataFig}
\vspace{-1em}
\end{figure*}

An instructive example of Client side transformations is the (randomized) \textit{$1$-deletion} depicted in Figure~\ref{dataFig}. 
This transformation takes in input a string $\file$ of $n$ elements, 
selects a component of $\file$ at random, say the $i$-th, and outputs 
the base $\file'$ of $n-1$ elements obtained from $\file$ by removing (deleting) the $i$-th element, and the 2-element deviation $ D$ consisting of the deleted value and its original position $i$ (in $\file$).

%deletion of one element in the string at a random position. 
%The outcome is a pair of strings where the first item is a \textit{basis} with one element less than the input, 
%and the second is a short \textit{deviation} that keeps track of the deleted value and its position in the original string. %(e.g., $(1,6)$). 
To build up intuition, the more \textit{deletions} a client performs before uploading its data the harder it is for a CSP to reconstruct the original data. %(what we call uncertainty). 
This increases the privacy of the outsourced data, however, the storage footprint on the client's side also increases. 
Section \ref{sec:transformations} elaborates on the transformations deployed in \name, while 
Section \ref{sec:analysis} analyzes the trade off between level of uncertainty and storage size on the client end.

\vspace{-3pt}
\subsection{Attacker Model}\label{sec:attacker_model}
We consider privacy against a computationally unbounded, honest-but-curious CSP, %denoted by \adversary. 
In detail, we assume this CSP %(a dishonest CSP) has 
knows the distribution of clients' raw files \distribution, 
reads all data outsourced by clients, and 
has unlimited computational power. 
The attacker's goal is to correctly reconstruct %(or guess)
 the clients' original files. 
We discuss how to measure the success probability of such attacks in the next section (\ref{sec:metrics}) through the `uncertainty metric'. 
Investigating how to reach security against a malicious attacker, either CSP or client is left as future work. 
%For the sole sake of privacy, we could consider a malicious CSP.
%However, the honest-but-curious assumption comes in hand to ensure the correctness of the system. 
%Indeed a malicious CSP could alter the uploaded data and prevent clients from retrieving the correct files. In this case, however, the clients can detect the unfair behaviour and stop using the system. 
%Therefore we consider such attacks out of scope. 

\subsection{Performance Metrics}\label{sec:metrics}
%In what follows, we define metrics to evaluate the compression capabilities and the privacy of our model. 
%We begin with the concepts related to the compression capabilities of our system, and end with the uncertainty metric to measure the privacy level. 

In what follows, \database denotes a database (collection of arbitrary files \file), 
\system denotes the dual deduplication system described in Section \ref{sec:model_system}, 
%\elena{generic, dual deduplication system, I think it's the first time we use this expression, but it feels it captures pretty well the intuition / flow of new approach}
\size\ is a function that takes as input a system \system, a party, e.g., \user or \cloud and a database \database, and returns
% a non-negative number corresponding to 
the size of the storage space required by the given party to store \database according to the system \system. 

Since our model describes systems where both \user and \cloud store some piece of information, 
it is natural to define three quantities to measure the system compression capability. 

\begin{description}
	\item[{Client Compression Ratio:}]
\begin{equation}\label{eq:ucr}
	\ucr = \frac{\size(\system, \user, \database)}{|\database|}.
\end{equation}

\item[{Cloud Compression Ratio:}]
	\begin{equation}\label{eq:ccr}
	\ccr = \frac{\size(\system, \cloud, \database)}{|\database|}.
\end{equation}

\item[{Global Compression Ratio:}]
	\begin{equation}\label{eq:gcr}
	\compratio  = \frac{\size(\system, \user, \database)+\size(\system, \cloud, \database)}{|\database|}.
	\end{equation}
\end{description}

Concretely, \compratio measures the compression capability of our system. 
The lower the value of \compratio the better the compression level and the smaller the overall storage space required. 
An ideal solution would have $\compratio< 1$. 
%In our case, sometimes we have  $\compratio> 1$; yet the local compression rates \ucr, \ccr\ are significantly smaller than $1$. This means that we have compression in \user and \cloud compared to storing the original data in either one of them.

Now we define a metric for evaluating the privacy of a system \system. 
The uncertainty metric \unmetric measures the degree of uncertainty a honest-but-curious CSP 
%(denoted as an attacker \adversary)
 faces when trying to retrieve clients' original files from the data they outsource. 
 To formally define \unmetric we need a distribution \distribution defined on the database \database. This essentially simulates the fact that CSP may know what are the most common files. 
Thus  we define 
\begin{description}
	\item[Uncertainty Metric:]
\begin{equation}\label{eq:uncertainty}
\unmetric(\file) = Prob_{\file\gets {\distribution}}[  \file^*\gets\adversary(\distribution, \file') | \file^*=\file]
\end{equation}
\end{description}
where $\file'$ is the outsourced data uploaded by the clients to the Cloud in correspondence to the original \file.
\section{\name}\label{sec:contribution}

We begin by describing the set of allowed transformations in \name. Then we explain the protocol in detail. 

% To allow compilation of the file
% !TeX root = ./../main.tex
\subsection{Allowed Transformations in \name}\label{sec:transformations}
From information theory, we know four functions to transform a string.
Namely, \begin{enumerate*}[label=(\roman*)]
	\item \label{lins}
 insert an element to a position, 
 \item \label{ldel}
 delete an element from a position, 
 \item \label{lswap}
  swap two elements, and 
 \item \label{lchange}
 change the value of a given position.
\end{enumerate*}
%functions to apply on the data to generate a number of possible transformations for our generalized deduplication approach.
%In order to have efficient deduplication, we also define some policies. In this section we take a look at the funtions and policies we use in our setup. 
%Throughout this work, we consider our data to be strings of a fixed length.
%we can use 4 functions to transform a string. These functions are as follows:
% \begin{enumerate}
% 	\item Insert an element to a position.
% 	\item Delete an element from a position.
% 	\item Swap two consecutive elements.
% 	\item Change the value of a position.
% \end{enumerate}
%The difference between these functions is 
Note that swap and change value do not change the length of the string, while insert and delete do. %operations change the length of the string.

\name allows \user and \cloud to determine policies on how to transform the data to obtain deduplication friendly strings. 
The aim is to minimize the number of operations to perform while achieving efficient deduplication rate on the Cloud and some level of privacy at the Client side. 
%Minimizing the number of operations increases the compression potential as the generated deviation needs to be stored to recover the original data.
%We provide policies on how to combine the aforementioned transformations %(denoted by $\mathcal{T}$) 
These policies typically require a metric to determine similarities among strings. 
A natural metric is 
%As mentioned in section~\ref{intro}, the deduplication procedure can be performed on string that are nearly identical. The similarity between the strings is measured by one of the following metrics.
the \textbf{Hamming distance}, indicating the number of positions with different values in two strings of the same length. 
This essentially tells us how many \emph{change value} operations we need to transform a string into another. \textbf{Swap Distance} indicates the number of operations to change a string into another using only the \emph{swap} and \emph{change value}.
\textbf{Damerau–Levenshtein Distance} is the most complete metric, essentially indicating the number of operations to transform one string into another if we use all 4 transformations \ref{lins}-\ref{lchange}~\cite{brill2000improved}.
% \begin{enumerate}
% 	\item \textbf{Hamming Distance}: The number of positions at which the corresponding symbols are different. This distance is only applicable for two strings of the same length. Hamming distance shows the number of operations we need to do to transform one string into another string if we only use the change value function.
% 	\item \textbf{Swap Distance}: The number of operations to transform one string to another one if we only use change value and swap functions.
% 	\item 
% 	\textbf{Damerau–Levenshtein distance}: The number of operations to transform one string into another if we use all 4 functions.
% \end{enumerate}

Previous work on GD focused primarily on changing values operations using Hamming or Reed-Solomon codes~\cite{rasmus,Vestergaard2019b}. 
%In these works, the notion of closeness is the Hamming Distance.
%In this work, 
Here we instead consider the three operations \ref{ldel}-\ref{lswap}-\ref{lchange}: delete (\deletions), swap (\swap) and change value (\changev). 
We discard the \emph{insert} function as it increases the size of a string, which is counterproductive for compressing the data.
%In previous works of the generalized deduplication, the hamming codes was used to ensure deduplication. In this sense, the original string is transformed to the basis with the lowest hamming distance. In this case, the bases were pre-defined~\cite{rasmus}. However, in this work, we use the swap distance in order to transform the generated bases into bases that can be deduplicated. The swap distance is more general and allows defining the bases with respect to data instead of using pre-defined bases.

In \name, 
%we use one transformation policy for \user and one for \cloud. The 
\user applies \deletions transformations on \database prior to sending data to \cloud to achieve the desired level of privacy.\footnote{Note that \deletions drops random components from the original string (Fig. \ref{dataFig}) and thus 
acts as a deletion channel~\cite{mitzenmacher2009} for the \cloud.
% (who gets the punctured string $\file'$). 
% Determining the original data chunk is a hard problem.
%, even if it could control the encoding of the original data before introducing it to the channel (which is not the case in our application).
The hardness of the reconstruction of the original data reflects into the \user's privacy level.}
\cloud applies \swap and \changev to reduce the distance between the strings uploaded by \user, generating strings that are suitable for genealized deduplication. 

\vspace{-4pt}
\subsection{Proposed Protocol}

%We propose the following algorithm for efficient deduplication:
%
%\begin{enumerate}
%	\item Define policies regarding the structure of already existing bases in the cloud. These policies is used by user to generate bases close to the existing bases.
%	\item Generate bases for original data using deletion operations. The operations are carried out until the bases of length $n_{b}$ are generated while satisfying the policies as much as possible.
%	\item Transform the bases in the cloud using swap and change value. In order to minimize the number of operation, the cloud finds the closest bases and transform one into another. For $\mathcal{N}_{b}$ bases, the swap distance between all bases is found by using the algorithm proposed by Amir, et al. in $O(\mathcal{N}_{b}n_{b}^{\frac{3}{2}}\log n_{b})$~\cite{amir}.
%	\item The bases are transformed, which will allow the deduplication to happen in the cloud.
%\end{enumerate}
%
%The bottleneck of this algorithm is the calculation of swap distances in step 3. In order to improve the efficiency of algorithm, we optimize this part by allowing a transformation of bases when their swap distance is lower than $T_{sd}$. Therefore, the expected number of calculations until we get a valid transformation is reduced and we do not need to calculate the swap distance between all pairs of strings.

We describe \name, our protocol for privacy-aware dual deduplication in multi client settings.
%There are two core subroutines in \name: 
%
%-\deletions: an element-deletion algorithm run on the client side that, given a string returns a shorter one and data on how the deletion happened. We denote the shortened string as basis and the information of the deletions as deviation. The deviation is used to recover the original data from the basis. An example can be seen in Fig.~\ref{fig:deletion}.
%
%- \swap: run by the cloud to make swap and change value changes to a new received string in order to obtain a target string that is already stored in the cloud. The output is a deviation indicating the changes and the ID of the target string.
%
%In detail, our system is made of the following algorithms:
The protocol is run between \user and \cloud, components of \system and is parametrized by
\begin{enumerate*}[label=(\alph*)]
	\item a distance metric \dist;
	\item a threshold value $\threshold>0$ that indicates the maximum number of operations allowed in the \cloud per string; 
\end{enumerate*} 
% a value for the size in number of symbols of $k$ bits of bases $n_b>0$;
% a (d) a limit on the maximum number of bases available in the \cloud $B$ that are used during the process of comparison using the \dist metric to decide on the best potential for compression. 
At initialization 
%\system holds an empty \database and 
\cloud holds an initial set $n_b$-size strings called bases $\baseset=\{\base_1,\ldots, \base_b\}$.
% for $b<B$. 
\footnote{We do not specify how to construct the initial set \baseset. However, the \compress algorithm of \name essentially allows to start from an empty $\baseset=\varnothing$ and populate it according to the uploaded files.}
$\baseset$ can be updated over time, but the full potential of updating \baseset  will be studied in future work.

At its core, the protocol performs a number of operations in the \user with a focus on privacy protection %and local deduplication (when possible) 
prior to uploading to the \cloud.
\cloud uses the information from each \user and the $\baseset$ to attempt deduplication of similar bases that are $\threshold$ operations away given a \dist metric.
If the data is similar to one basis in the $\baseset$, it will be deduplicated, otherwise, it will be stored as it is.
In the following, we provide a description of the various operations of \name and where they take place.

\begin{description}
\item[$\setup(\baseset)$:] 
	This algorithm is run by the \cloud periodically. 
	It takes as input a set of bases \baseset
	% where each base has size $|\base|=\sbase$. 
	and outputs a policy \policy that concisely describes \baseset.\footnote{Policies ensure efficient deduplication according to the current set of bases. 
	To give an example \policy could encode 
	`binary strings of length $n_b$ with $5$ number of 0s towards the end'.
}
	
%	\item[$\deletions(\policy,\file)$:] 
\item[$\upload(\policy,\file)$:] 
	This algorithm is run by \user using a given \policy
	On input of a file \file, \deletions are applied to \file according to \policy 
	until it results in a string \outsource with the size of $n_{b}$, 
	and its corresponding deviation \local. 
	Then, it generates a unique file identifier \id for the pair (\outsource, \local). 
	It outputs the pair (\id,\outsource) to the CSP, 
	while the pair (\id,\local) are stored locally on the \user.

	\item[$\compress(\baseset, \id, \outsource)$:] 
	This algorithm is run by the \cloud. 
	It takes as input \baseset, 
	%the current database \database, 
	a file identifier \id and a string \outsource. 
	Then, it performs the following steps:
\vspace{.3em}

	1. If $\outsource =\hit{\base}\in \baseset$ set $\deviation=\varnothing$ and go to step 4.
\vspace{.3em}
	
	2. If $\exists \hit{\base}\in \baseset$ that $0<\dist( \outsource, \baseset) \leq\threshold$ %--for the system's distance metric \dist and system's threshold value \threshold-- 
	then: 

%\hspace{1em}- Find the element $\hit{\base}$ in  \baseset that is closest to \outsource.

\hspace{1em}- Repeatedly apply \swap and \changev on \outsource until reaching the target $\hit\base$. Denote by $\deviation$ the corresponding string of deviations. 
\vspace{.3em}

	3. If $\forall\hit{\base}\in\baseset$ that $\dist(\outsource,\baseset)>\threshold$ set $\hit{\base}:=\outsource$, 
	set $\deviation=\varnothing$ and add \hit{\base} to \baseset. 
\vspace{.3em}
	
	4. Perform deduplication by storing %in \database 
	the file identifier \id, a pointer to \hit{\base} and the string \deviation. 
	
	\item[$\get(\id, \local, \cdot)$:]
	%This algorithm is actually an interactive protocol between \user and \cloud. 
	This algorithm is initiated by \user sending an \id, symbolizing a request to retrieve the item that was outsourced with that \id. 
	Upon receiving a response \outsource from \cloud, the
	\user uses the information encoded in the local deviation \local connected to \id to invert the deletions that led to \outsource, thus reconstructing \file.

      \item[$\decompress(\database, \cdot)$:]
      %	This algorithm is actually an interactive protocol between \user and \cloud.
      This algorithm is run on \cloud upon receiving a query of the form \id.
      	\cloud checks whether it has stored \id. 
      	If not, it ignores the query, otherwise it retrieve the corresponding item (\id, \base,\deviation) and inverts the generalized deduplication performed by \compress  to reconstruct the decompressed \outsource corresponding to the outsourced string. \outsource is sent back to \user.		 
\end{description}

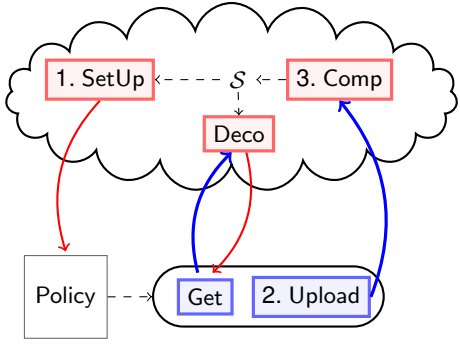
\begin{figure}[!t]
	\centering
	\begin{tikzpicture}[font=\small\sffamily, thick, scale=0.7]
	
	\node[cloud, cloud puffs=19, cloud ignores aspect, minimum width=6cm, minimum height=2.5cm, align=center, draw] at (0cm, 0cm) {};
	\node [alias=cloud] at (0,-1.2){};

		\node [alias=setup,style={rectangle, draw=red!60, fill=red!5, very thick}] at (-2.5,0.33){1. \setup};

		\node [alias=set] at ([xshift=45pt] setup.east){$\baseset$};

		\node [alias=swap,style={rectangle, draw=red!60, fill=red!5, very thick}] at ([xshift=45pt] set.east){3. \compress};
		
		\node [alias=rec,style={rectangle, draw=red!60, fill=red!5, very thick}] at (0.14,-.7){\decompress};
		
%%%% CLIENT	
		
		\draw (.7,-3.2) node[anchor=north,rounded rectangle,minimum width=3.3cm, minimum height=.8cm,
		draw, alias=one]{}; 

		\node[alias=get,style={rectangle, draw=blue!60, fill=blue!5, very thick}] at (-.5,-3.8) {{\get}};
		
		\node[alias=upload,style={rectangle, draw=blue!60, fill=blue!5, very thick}] at ([xshift=42pt]get.east) {2. \upload};
		
		\node [alias=policy, style={rectangle,  minimum width=1.1cm,minimum height=1.1cm, draw=black!60, thin}] at ([xshift=-60pt] get.west){$\policy$};
		%	\node[alias=del,style={rectangle, draw=blue!60, fill=blue!5, very thick}] at ([xshift=20,yshift=-13pt]one) {{2. \deletions}};
		
		%	\path[<->,very thick] (one) edge [bend right] ([xshift=-6pt] cloud);

	\path[->, dashed,thin] (policy) edge (one);
	\path[->, thick,red] ( setup.south) edge [bend right] ([yshift=2pt]policy.north);
		
	\path[->, very thick,blue] ( upload.east) edge [bend right] (swap.south);
	
		\path[->, very thick,blue] ( [xshift=-4pt,yshift=3pt] get.north) edge [bend left] ([xshift=-4pt]rec.south);
	
		\path[->, thick,red] ( [xshift=4pt] rec.south) edge [bend left] ([xshift=4pt,yshift=3pt]get.north) ;
		
			\path[->, dashed,thin] (set.west) edge (setup.east);
			\path[->, dashed,thin] (swap.west) edge (set.east);
			\path[->, dashed,thin] ([yshift=3pt]set.south) edge (rec.north);

		%%%%%%%%%%%%%%%%%%5
%		\node [alias=file, style={rectangle,  minimum width=3cm, draw=black!60, thin}] at (5.6,-5.5){file};
%	
%		\node [alias=a, style={rectangle,  minimum width=2.3cm, draw=black!60, thin}] at ([yshift=40pt,xshift=-18pt]file){\phantom{f}};
%		\node [alias=b, style={rectangle,  minimum width=.8cm, draw=black!60, fill=blue!30, thin}] at ([xshift=18pt]a.east){\phantom{f}};
%		\node [alias=c, style={rectangle,  minimum width=2.3cm, draw=black!60, thin}] at ([yshift=40pt,xshift=-10pt]a){\phantom{f}};
%		\node [alias=d, style={rectangle,  minimum width=1.1cm, draw=black!60, fill=red!30, thin}] at ([xshift=23pt]c.east){\phantom{f}};
%		\node [alias=base, style={rectangle,  minimum width=.9cm, draw=black!60, fill=red!30, thin}] at ([yshift=40pt]c){\phantom{f}};
%	
%	\draw[->, dashed,thin] (file.north) -> ([yshift=20] file.north) node[midway, right] {\deletions};
%	\draw[->, dashed,thin] (a.north) -> ([yshift=20] a.north) node[midway, right] {\swap};
%		\draw[->, dashed,thin] (c.north) -> ([yshift=20] c.north) node[midway, right] {\gdd};
	
	\end{tikzpicture}
	\vspace{-.8em}
	\caption{\name System Model for Secure, Multi-client Dual Deduplication}
	\label{fig:deletion}
	\vspace{-1.5em}
	
\end{figure}

\vspace{-3.7pt}
% To allow compilation of the file
% !TeX root = ./../main.tex
\section{Analysis of Yggdrasil}\label{sec:analysis}

% To improve readability, the notation is collected in Table \ref{tab:notation}.

In the following, we analyze the compression rate and the privacy (uncertainty measure) achieved by \name. In this section, unless stated otherwise, we use $\log(x)$ as the logarithm in base $2$ of $x$. 
\vspace{-5pt}
\subsection{Client Compression Ratio}
We begin our analysis with studying the compression ratio on the client side, i.e., \ucr. 
%Recall that in \name, the original data is made of strings $\file \in (\{0,1\}^k)^{n_o}$, i.e., vectors of $n_o$ entries, each entry being $k$ bits long. 
%Prior to upload, \user transforms \file into a shorter and fix-length string $\outsource\in (\{0,1\}^k)^{n_b}$ for  \cloud; and a local deviation \local that is stored on the client machine. 
To compute  \ucr for one string \file, 
we need to accurately measure the size of \local. 
Assuming \user performed $\nodel=n_o-n_b$ subsequent \deletions on \file, 
then \local contains the \nodel deleted values (each value has $k$ bits) 
and a pointer to their original locations in \file (each pointer has $\lceil\log(n_{o})\rceil$).
Therefore, the required storage needed in \user is equal to:
\begin{equation*}
\size(\system, \user, \file) = \nodel(\lceil \log(n_{o})\rceil + k) + \sizefid,
\end{equation*}
where $\sizefid$ is the size in bits of a file identifier ($\sizefid=\size(\id)$).
If our \database has $f$ files, the required storage on the \user side for the whole \database is
\begin{equation*}
\size(\system, \user, \database) = f(\nodel(\lceil \log(n_{o})\rceil + k) + \sizefid).
\end{equation*}

If \database contains \nof strings, the total storage size required for \database prior to the transformations of \name is
\begin{equation*}
|\database| = \nof\cdot k\cdot n_{o}.
\end{equation*}
%For each deleted item, \local contains (at most\footnote{
%Notably, there are several ways to reduce the footprint of the 'pointers' in \local (e.g., using deduplication, storing difference from previous pointer...).}) 
%one $k$-bit element (representing the original value) 
%and a $\log(n_o)$-bit pointer to its original position in \file. 
%In addition, \user needs to also store the file identifier \id to be able to retrieve the outsourced data item, 
% This equation leads to the following results for the compression ratio on \user, we have:
%%	$$s_{b} = s_{o} - x\cdot k,$$
%\begin{equation*}
%\frac{\size(\system, \user, \file)}{\size(F)} = \frac{x\cdot (\lceil \log(n_{o})\rceil + k) + \sizefid}{k\cdot n_o}.
%\end{equation*}

Thus,
\begin{equation}
%\begin{split}
\ucr = \frac{\nodel(\lceil \log(n_{o})\rceil + k) + \sizefid}{k\cdot n_o}.
%\end{split}
\label{eq:ygg_ucr}
\end{equation}
 
%We remark that  \eqref{eq:ygg_ucr} holds for a single data item \file as well as for the whole database \database. 
%Indeed, the size of the (uncompressed) database is simply $\size(\database)=(\#~records) \cdot \size(\file) = \nof \cdot n_o$. 
%In \name, \user stores one pair (\id, \local) per record, thus there are \nof such local deviation stored on \user machine. 
%Hence the ratio in Equation \ref{eq:ygg_ucr} remains unchained (both numerator and denominator get multiply by the same factor \nof). 

Clearly $\ucr<1$ if and only if
$$\sizefid + x\cdot(\lceil\log(n_{o})\rceil + k) < k\cdot n_{o}.$$
%Table \ref{tab1} collects numerical estimates of \ucr for various values of $n_o$ and $k$. Notably, Table \ref{tab:M1} shows that for many realistic values $\ucr<1$ and remarkably reaches rates as low as $0.1159$. 

\subsection{Cloud Compression Ratio}
%We recall that 
The data stored in \cloud The size of data stored in the \cloud consists of $b$ basis, where each basis has $n_{b}$ symbols of size $k$; one file identifier \id per string $\file$ and the \deviation generated in \compress procedure for deduplicated strings. Each \swap in \cloud adds $2\lceil\log(n_{o})\rceil$ and each \changev adds $k + \lceil\log(n_{o})\rceil$ to \deviation. As there are $\nof - \nob$ deduplicated strings and the number of operations in \cloud for each \file is bounded by \threshold, we have
\begin{equation*}
\size(\system, \cloud, \database) \leq \nob \cdot k\cdot n_{b} + \nof\cdot \sizefid + (\nof - \nob)\cdot \threshold(2\lceil\log(n_{o})\rceil).
\end{equation*}
 Therefore, for the compression ratio in \cloud, we have:
\begin{equation*}
\ccr = \frac{\sizefid + 2\threshold\lceil\log(n_{o})\rceil + \baseF( k\cdot n_{b} - 2\threshold\lceil\log(n_{o})\rceil)}{k\cdot n_{o}}.
\end{equation*}
where $\baseF$ is the fraction of number of bases to the number of original strings, i.e., $r=\frac{\nob}{\nof}$
%$$n_{b} = n_{o}-x,$$
%Since the maximum number of possible $1$-deletions is equal to $n_{o} - 1$, then:
%$$x \leq n_{o} - 1.$$
The condition for achieving a compression ratio of less than one is:
$$r \leq 1 - \frac{\sizefid - k\cdot \nodel}{k\cdot (n_{o}-\nodel)- 2\threshold\lceil\log(n_{o})\rceil}\Leftrightarrow \ccr \leq 1.$$

%Considering the overhead of using fingerprints in the cloud and client, there is an extra storage size of $s_{d}$ in both client and cloud. Hence, the client and cloud compression ratios become:
%$$$$  
\subsection{Global Compression Ratio}
The global compression ratio of the system is given by the sum of the \user compression ratio and the \cloud one. Thus, 
\begin{equation}
\footnotesize
\compratio = \frac{2\sizefid + k\cdot x + (2\threshold + \nodel)\lceil\log(n_{o})\rceil + \baseF( k\cdot n_{b} - 2\threshold\lceil\log(n_{o})\rceil)}{k\cdot n_{o}}.
\label{eq:C1}
\end{equation}

\subsection{Uncertainty of Multiple $1$-Deletions}

%The number of possible bases which we hit from a string is equal to:
%\begin{equation*}
%\rmetric = {n_{o}\choose x}.
%\end{equation*}

%The uncertainty metric, i.e., the number of strings that can be generated by a basis is equal to the number of strings that has the provided basis as a subsequence. This value is equal to:

We now calculate the uncertainty of a data item after \user performs $\nodel$ 1-deletions. 
We consider the probability distribution \distribution over the set of $k$-bit symbols to be uniformly random, i.e., every symbol %(or component of \file) 
has the same probability $1/2^k$ to be selected. 
In this setting, our definition of uncertainty (Eq.~\eqref{eq:uncertainty}) states that the uncertainty of a string $\unmetric(\outsource)$ is equal to $1$ over the number of original strings $\tilde{\file}$ that can be generated by the base $\outsource$ output to the \cloud.
% by $\upload(\policy,\file)$. 
Let \nopreim denote this value, then~\cite{bartle2000introduction}:%\footnote{The proof for this claim is rather long and involved and has been omitted to meet the page limit. The detailed proof is for review on the full version of this paper available at XXXXX.}

\begin{equation}
\nopreim = \sum_{j=0}^{n_{o}-n_{b}} {n_{o} \choose j+n_{b}} (2^{k}-1)^{n_{o}-n_{b}-j}.
\label{eq:nopreim}
\end{equation}
Intuitively, \nopreim counts the number of `preimagines of \upload', i.e., how many $n_o$-element strings \file  can generate the same base \outsource for a combination of $\nodel$ \deletions.
For large enough $n_o$ and $n_b \geq n_b$, a good lower bound is to consider the first term in the summation, i.e.,  $\nopreim \geq {n_{o} \choose n_{b}} (2^{k}-1)^{n_{o}-n_{b}}.$ 
%This equation leads to the following result:
%\begin{equation*}
%\mathcal{N} = {n_{o}\choose n_{b}}(2^{k}-1)^{n_{o}-n_{b}} \pFq{2}{1}{1, n_{b} - n_{o}}{ n_{b} + 1} {\frac{1}{1-2^{k}}},
%\end{equation*}
%
%where $\pFq{2}{1}{i, j}{k}{l}$ is the hypergeometric function. 
Using Eq.~\ref{eq:nopreim}, the uncertainty metric for a record $\file\gets\distribution$ is
%\begin{equation}
%\unmetric = \frac{(2^{k}-1)^{n_{b}-n_{o}}}{{n_{o}\choose n_{b}} \pFq{2}{1}{1, n_{b} - n_{o}}{ n_{b} + 1} {\frac{1}{1-2^{k}}}}.
%\end{equation}
\begin{equation*}
\unmetric(\file) = \frac{1}{\nopreim} =  \frac{(2^{k}-1)^{n_{b}-n_{o}}}{\sum_{j=0}^{n_{o}-n_{b}} {n_{o} \choose j+n_{b}} (2^{k}-1)^{-j}.}.
\end{equation*}
Table~\ref{tab:M1} shows this number for various symbol sizes $k$, original string sizes $n_{o}$, and basis size $n_{b}$.
Even for small sequences of $n_o = 15$ with $k = 2$, the uncertainty is in the order of $10^{-7}$.
For more realistic cases, e.g., $n_o = 1000$, $n_b = 500$ and $k = 8$, the uncertainty metric is $10^{-1503}$, creating a high degree of potential uncertainty on \cloud.

\begin{table}[!t]
	\caption{Numerical computation of \nopreim and \unmetric for varying $k,n_o$ and $n_b$. }
	\vspace{-.8em}
	\centering
	\begin{tabular}{|c|c|c||c|c|}
		\hline
		$k$ & $n_{b}$ & $n_{o}$ & $\nopreim$ & $\unmetric$ \\ 
		\hline
		\rule{0pt}{8pt}2 & 10 & 15 & $8.53\times 10^{6}$ & $1.17\times 10^{-7}$ \\
		\hline
		\rule{0pt}{8pt}4 & 10 & 15 & $2.35\times 10^{9}$& $4.26\times 10^{-10}$ \\
		\hline
		\rule{0pt}{8pt}8 & 10 & 15 & $3.24\times 10^{15}$& $3.08\times 10^{-16}$\\
		\hline
		\rule{0pt}{8pt}2 & 100 & 150 & $1.72\times 10^{64}$& $5.81\times 10^{-65}$\\
		\hline
		\rule{0pt}{8pt}4 & 100 & 150 & $1.32\times 10^{99}$ & $7.58\times 10^{-100}$\\
		\hline
		\rule{0pt}{8pt}8 & 100 & 150 & $4.28\times 10^{160}$& $2.34\times 10^{-161}$\\
		\hline
		\rule{0pt}{8pt}2 & 500 & 1000 & $1.47\times 10^{538}$& $6.80\times 10^{-539}$\\
		\hline
		\rule{0pt}{8pt}4 & 500 & 1000 & $3.21\times 10^{887}$& $3.12\times 10^{-888}$\\
		\hline
		\rule{0pt}{8pt}8 & 500 & 1000 & $5.05\times 10^{1502}$& $1.98\times 10^{-1503}$\\
		\hline
	\end{tabular}
\vspace{-1em}
	\label{tab:M1}
\end{table}

\subsection{Most Probable String}

%In this section, we address the problem of most probable string. This problem can be formulated as follows:

Let $P(\file=o|\outsource,\distribution)$ denote the probability of an original string $o$ of length $n_o$, given a basis \outsource and a probability distribution of \distribution for the symbols in the original strings.
From an attacker's perspective, the key is to identify the most probable string, i.e., the string $o$ in the set of all strings of size $n_{o}$, that has 
$\max\limits_{o} P(\file=o|\outsource,\distribution)$.
From a system designer's perspective, a similar question is interesting to achieve a higher privacy in the system.
Namely, the system wants to \emph{minimize} this probability.
% We suppose that all deletions are 1-deletion. This assumption does not reduce the generality of the calculations as any $y$-deletion can be mapped into $y$ 1-deletions. 

In order to generate the original data, the attacker needs to add $n_{o} - n_{b}$ symbols to the basis in arbitrary positions. 
%It is clear that, The most probable string is the string that is generated by more than 1 set of functions, i.e., the duplicates. The most number of duplicates occur when we insert a value $i$ between the elements in basis that already has value $i$. For example in the following Basis:
%$$[2,2,1,1,1,2,0,3,1,1]$$
%If we insert 1 before the positions 2, 3, 4 and 5, all the results are duplicates, which is the highest number of possible duplicates. Also, if we insert 2 in positions 0, 1 and 2, all the results are duplicates. 
In this setup, The most probable strings is the string with the most number of duplicates in the reconstruction. The duplicates occur when we insert a value $i$ between consecutive elements in basis that already has value $i$. 
Suppose the longest consecutive elements of value $i$ in \outsource has length $l_{i}$, hence, the number of possible duplicates is equal to:
$$\max\limits_{i}\sum_{j=0}^{n_{o}-n_{b}} (l_{i}+j) = \max\limits_{i}\frac{1}{2} (n_{o}-n_{b}+1)(2\cdot l_{i} +n_{o}-n_{b}).$$

Therefore, the most probable string has a probability of:
\begin{equation}
\begin{split}
&\max\limits_{o}P(\file=o|\outsource,\distribution) = \\
& \max\limits_{i}\Big(\frac{1}{2} p_{i}^{n_{o}-n_{b}}\cdot (n_{o}-n_{b}+1)(2\cdot l_{i} +n_{o}-n_{b})\Big),
\end{split}
\end{equation}
%where $l_{i}$ is the length of longest consecutive symbols of value $i$ in \outsource.
Clearly, there is a strong dependence on the probability distribution of the original strings.
%The string with this probability is generated by extending the size of the longest consecutive elements of the value with highest probability until the size of the string is equal to $n_{o}$.
To reduce this probability for a set of data, we can define several policies:
%\begin{itemize}
  
\textbf{Policy 1:} Set the probability distribution of the elements in the basis to be as close as possible to uniformly random distribution, e.g., a ciphertext. This may be counterproductive for the compression process. However, approaching a uniform distribution provides clear privacy advantages. A system designer can try to select the level of protection desired.

\textbf{Policy 2:} A basis does not include consecutive identical symbols, especially the symbols with higher probability.

          % \end{itemize}

% To allow compilation of the file
% !TeX root = ./../main.tex
\noindent
\begin{figure*}[!b]
	\centering
	\subfloat[$k=2$, $n_{b} = 950$]{\begin{tikzpicture}
		[scale=0.6]
		\begin{axis}[
		xlabel= $\threshold$,
		ylabel=$Compression\; Rate$,
		xlabel style={font=\large},
		grid=major,
		legend pos= outer north east]
		
		\addplot[mark= +, smooth ,black, thick] plot coordinates {
			(5, 1.13) %1
			(15, 1.21) %2
			(25, 1.27)%3
			(35, 1.32)%4
			(45, 1.27)%5
			(55, 1.17)%6
			(65, 1.12)%7
			(75, 1.09)%8
			(85, 0.94)%9
			(95, 0.81)%10
			(105, 0.62)%11
			(115, 0.55)%MIN
			(125, 0.58)%13
			(135, 0.61)%14
			(145, 0.64)%15
			(155, 0.68)%16
			(165, 0.71) %1
			(175, 0.76) %2
			(185, 0.8)%3
			(195, 0.86)%4
			(205, 0.92)%5
			(215, 0.98)%6
			(225, 1.09)%7
			(235, 1.16)%8
			(245, 1.21)%9
			(255, 1.27)%10
			(265, 1.33)%11
			(275, 1.45)%12
		};
		\addlegendentry{$\compratio$}
		
		\addplot[mark = |, smooth, red, thick] plot coordinates {
			(5, 0.23) %1
			(15, 0.23) %2
			(25, 0.23)%3
			(35, 0.23)%4
			(45, 0.23)%5
			(55, 0.23)%6
			(65, 0.23)%7
			(75, 0.23)%8
			(85, 0.23)%9
			(95, 0.23)%10
			(105, 0.23)%11
			(115, 0.23)%12
			(125, 0.23)%13
			(135, 0.23)%14
			(145, 0.23)%15
			(155, 0.23)%16
			(165, 0.23) %1
			(175, 0.23) %2
			(185, 0.23)%3
			(195, 0.23)%4
			(205, 0.23)%5
			(215, 0.23)%6
			(225, 0.23)%7
			(235, 0.23)%8
			(245, 0.23)%9
			(255, 0.23)%10
			(265, 0.23)%11
			(275, 0.23)%12
			
		};
		\addlegendentry{$\ucr$}

		\addplot[mark=x, smooth,blue, thick] plot coordinates {
			(5, 0.9) %1
			(15, 0.98) %2
			(25, 1.04)%3
			(35, 1.09)%4
			(45, 1.04)%5
			(55, 0.94)%6
			(65, 0.89)%7
			(75, 0.86)%8
			(85, 0.71)%9
			(95, 0.58)%10
			(105, 0.39)%11
			(115, 0.32)%MIN
			(125, 0.35)%13
			(135, 0.38)%14
			(145, 0.41)%15
			(155, 0.45)%16
			(165, 0.48) %1
			(175, 0.53) %2
			(185, 0.57)%3
			(195, 0.63)%4
			(205, 0.69)%5
			(215, 0.75)%6
			(225, 0.86)%7
			(235, 0.93)%8
			(245, 0.98)%9
			(255, 1.04)%10
			(265, 1.08)%11
			(275, 1.22)%12
		};
		\addlegendentry{$\ccr$}

		\end{axis}
		\end{tikzpicture}}	
	\subfloat[$k=2$, $n_{b} = 900$]{\begin{tikzpicture}
		[scale=0.6]
		\begin{axis}[
		xlabel= $\threshold$,
		ylabel=$Compression\; Rate$,
		xlabel style={font=\large},
		grid=major,
		legend pos= outer north east]
		
		\addplot[mark= +, smooth ,black, thick] plot coordinates {
			(5, 1.31) %1
			(15, 1.39) %2
			(25, 1.28)%3
			(35, 1.22)%4
			(45, 1.11)%5
			(55, 1.02)%6
			(65, 0.89)%7
			(75, 0.74)%8
			(85, 0.65)%MIN
			(95, 0.71)%10
			(105, 0.77)%11
			(115, 0.82)%12
			(125, 0.86)%13
			(135, 0.91)%14
			(145, 0.97)%15
			(155, 1.03)%16
			(165, 1.09) %1
			(175, 1.15) %2
			(185, 1.21)%3
			(195, 1.28)%4
			(205, 1.34)%5
			(215, 1.41)%6
			(225, 1.48)%7
			(235, 1.55)%8
			(245, 1.59)%9
			(255, 1.65)%10
			(265, 1.71)%11
			(275, 1.78)%12
		};
		\addlegendentry{$\compratio$}
		
		\addplot[mark = |, smooth, red, thick] plot coordinates {
			(15, 0.45) %2
			(25, 0.45)%3
			(35, 0.45)%4
			(45, 0.45)%5
			(55, 0.45)%6
			(65, 0.45)%7
			(75, 0.45)%8
			(85, 0.45)%9
			(95, 0.45)%10
			(105, 0.45)%11
			(115, 0.45)%12
			(125, 0.45)%13
			(135, 0.45)%14
			(145, 0.45)%15
			(155, 0.45)%16
			(165, 0.45) %1
			(175, 0.45) %2
			(185, 0.45)%3
			(195, 0.45)%4
			(205, 0.45)%5
			(215, 0.45)%6
			(225, 0.45)%7
			(235, 0.45)%8
			(245, 0.45)%9
			(255, 0.45)%10
			(265, 0.45)%11
			(275, 0.45)%12
			
		};
		\addlegendentry{$\ucr$}

		\addplot[mark=x, smooth,blue, thick] plot coordinates {
			(5, 0.86) %1
			(15, 0.95) %2
			(25, 0.83)%3
			(35, 0.77)%4
			(45, 0.66)%5
			(55, 0.57)%6
			(65, 0.44)%7
			(75, 0.29)%8
			(85, 0.2)%MIN
			(95, 0.26)%10
			(105, 0.32)%11
			(115, 0.37)%12
			(125, 0.41)%13
			(135, 0.46)%14
			(145, 0.52)%15
			(155, 0.58)%16
			(165, 0.64) %1
			(175, 0.7) %2
			(185, 0.76)%3
			(195, 0.83)%4
			(205, 0.89)%5
			(215, 0.96)%6
			(225, 1.03)%7
			(235, 1.1)%8
			(245, 1.14)%9
			(255, 1.2)%10
			(265, 1.26)%11
			(275, 1.33)%12
		};
		\addlegendentry{$\ccr$}

		\end{axis}
		\end{tikzpicture}}		
	\subfloat[Global Compression Ratio.]{\begin{tikzpicture}
		[scale=0.6]
		\begin{axis}[
		xlabel= $n_{b}$,
		ylabel=$\compratio$,
		xlabel style={font=\large},
		grid=major,
		legend pos= north east]
		
		\addplot[mark=+, smooth ,black, thick] plot coordinates {
			(800,0.7924)
			(825,0.7363)
			(850,0.6894)
			(875,0.6243)
			(900,0.5527)%MIN
			(925,0.6118)
			(950,0.6583)
			(975,0.7137)
			(1000,0.8322)
		};
		\addlegendentry{$k=2$}
		
		\addplot[mark=x, smooth, dotted, red, thick, mark options={solid}] plot coordinates {
			(800,0.6933)
			(825,0.6718)
			(850,0.6332)
			(875,0.5919)
			(900,0.5341)
			(925,0.5118)
			(950,0.5927)
			(975,0.6818)
			(1000,0.7412)
			
		};
		\addlegendentry{$k=4$}

		\addplot[mark =o, smooth,green, mark options={solid}] plot coordinates {
			(800,0.7241)
			(825,0.6819)
			(850,0.6332)
			(875,0.5833)
			(900,0.5419)
			(925,0.5133)
			(950,0.4912)%MIN
			(975,0.5418)
			(1000,0.5974)
		};
		\addlegendentry{$k=8$}
		
		\addplot[mark = oplus, smooth, blue, thick] plot coordinates {
			(800,1.3722)
			(825,1.1844)
			(850,1.0268)
			(875,0.9781)
			(900,0.9329)
			(925,0.9117)
			(950,0.8922)
			(975,0.8332) %MIN
			(1000,0.8818)
		};
		\addlegendentry{$k=16$}
		
		\end{axis}
		\end{tikzpicture}	}	
	
	\caption[2]{Global, Client and Server Compression Rates. 
		The size of original records is fixed to $\size(\file)=n_{o} = 1024$ bits.}
	\label{fig:result1}
\end{figure*}
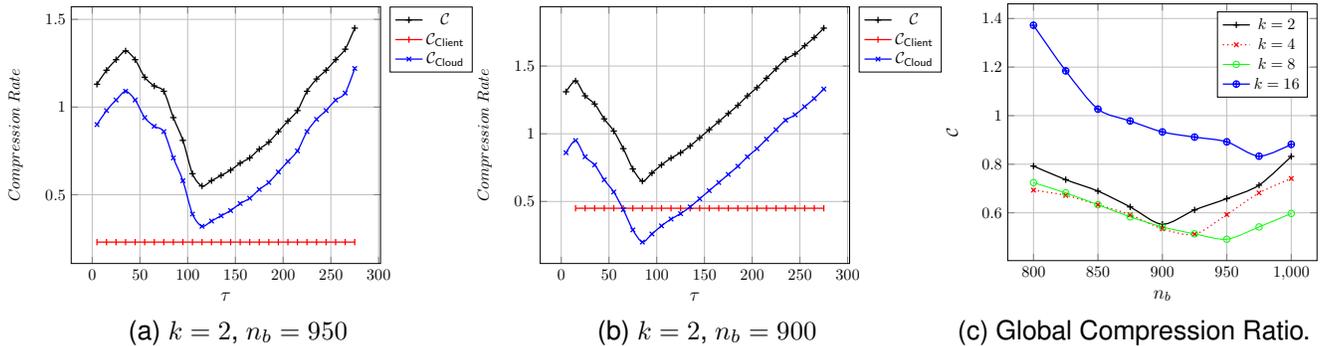

\section{Simulation Results and Discussion}\label{sec:experiments}

In this section, we first show our simulation results and then discuss the results and the performance of \name. 
In order to validate our system, we use the data of $18$~GB of HDFS logs as our dataset.
We developed a C++ implementation of a \name client applying random $1$-deletions and a server performing swap and change value operations.
The client and server both carry out deduplication steps separately.
For the sake of storage friendly implementation, we defined the \id of bases to be a global variable auto-incremented by the Cloud.
This means that each \id has a storage size of $\log(\mathcal{N}_{b})$ where $\mathcal{N}_{b}$ is the number of bases after deduplication.
This can be later changed to standard fingerprint functions, e.g., SHA-1, SHA-256, in the Cloud.
The impact of these slightly larger \id is minor for bases of 1~KB of more in size, which is the case we focus on in the following.  
%Te the client performs deletion operations and sends the data to the cloud. The cloud performs swap and change value operation in order to deduplicate as much as possible. 

Fig.~\ref{fig:result1} shows %the  compression rate of the client and the cloud as well as the total compression rate of the system.
that a judicious selection of the number of transformations results in a total compression ratio $\compratio = 0.5527$ for $n_{b} = 950$ and $\compratio = 0.6583$ for $n_{b} = 900$.
For a fixed value of $n_{b}$, the compression ratio in the \user is constant as the client always performs $n_{o} - n_{b}$ deletions.
However, the compression ratio of the \cloud differs depending on the maximum number of allowed swap and change value operations ($\threshold$).
A small $\threshold$ results in high compression rates. In fact, overall stored data may be higher than the original size because the number of bases that are deduplicated is low.
Thus, the extra storage required to store the operations is larger than the gains from deduplication.
A higher $\threshold$ allows for more bases to be deduplicated.
We show that there exists a minimum point for both \cloud and overall compression ratio.
This is the best compression ratio achievable for a given pair ($k$, $n_{b}$) for $n_o = 1024$.
Our analysis shows that this minimum is close to the median of the swap distance between all bases, which can be used in future work as a heuristic for the optimal value of $\threshold$.
Larger $\threshold$ values result in an increase in the compression ratio, because there are diminishing returns on the deduplication potential and the cost of recording additional operations results in much higher storage costs.

Fig.~\ref{fig:result1}.(c) shows the best compression ratio achieved for various values of $n_{b}$ and $k$ and a fixed $n_o$ and shows an optimal selection of $n_b$ and $k$ to achieve the highest overall compression.
%Independently of $k$, we observe that initially \compratio decreases with increasing $n_{b}$. 
%However, from a certain point on the compression rate increases linearly in $n_b$.
%Irrespective of the value of $k$, we show that this optimal  that increasing the value of $n_{b}$ initially reduces the number of 1-\deletions. % but results in higher deduplication cost on \cloud.
%However, when the size of the basis becomes `too large', the cloud will need to carry out a large number of swap and change value operations in order to perform deduplication.
%The overhead of storing these operations in the deviation results in the increase in the compression rate.
Fig.~\ref{fig:result1}.(c) shows that this minimum point is reached for a lower $n_b$ when $k$ is smaller, e.g., the best compression ratio for $k=2$ is achieved at $n_{b} = 900$, while the best compression ratio for $k=8$ is achieved at $n_{b} =950$.
%Note that increasing $k$ until $k=8$ results in a better compression ratio.
% This behavior is  because of the fact that the size of deviation is reduced with respect to the size of symbols.
We also observe that $k=8$ provides the best potential for compression overall ($49$~\% of original data) with only small degradations of the compression rate around the optimal point.
This is important in practice as the \user's may aim to achieve different uncertainty - storage trade-offs and $k=8$ corresponds to byte representations that are good to achieve efficient software implementations.
Note that for $(k,n_b,n_o)=(8,950,1024)$ the number of possible original strings that could generate each basis uploaded to the \cloud is around $10^{293}$.
%This is explained by the relation between the size of deviation and size of symbols.
%In $k=2$, each elimination adds a block of size $s_{d}=12$ to the deviation, which is $6 \times k$, while $k=8$ each elimination adds $s_{d} = 18$, which is less than $3\times k$.
For $k = 16, 32$ the cost of storing the deviation for each elimination in the \user is high, which limits the overall potential for compression in the \user and, thus, in the overall system.
%However, this trend changes for $k=16, 32$, because deviation growth per elimination is too high, reducing the potential to carry out a large number of elimination, thus, reducing the deduplication potential in the client.
Table~\ref{tab1} provides details for the \cloud and \user compression ratios for different $k$ and $n_b$.

%\noindent
%\begin{figure}[!t]
%	\centering
%\begin{tikzpicture}
%		[scale=0.6]
%		\begin{axis}[
%		xlabel= $n_{b}$,
%		ylabel=$\compratio$,
%		xlabel style={font=\large},
%		grid=major,
%		legend pos= outer north east]
%		
%		\addplot[mark=+, smooth ,black, thick] plot coordinates {
%		(800,0.7924)
%		(825,0.7363)
%		(850,0.6894)
%		(875,0.6243)
%		(900,0.5527)%MIN
%		(925,0.6118)
%		(950,0.6583)
%		(975,0.7137)
%		(1000,0.8322)
%		};
%		\addlegendentry{$k=2$}
%		
%		\addplot[mark=x, smooth, dotted, red, thick, mark options={solid}] plot coordinates {
%		(800,0.6933)
%		(825,0.6718)
%		(850,0.6332)
%		(875,0.5919)
%		(900,0.5341)
%		(925,0.5118)
%		(950,0.5927)
%		(975,0.6818)
%		(1000,0.7412)
%			
%		};
%		\addlegendentry{$k=4$}
%		
%		
%		\addplot[mark =o, smooth,green, mark options={solid}] plot coordinates {
%		(800,0.7241)
%		(825,0.6819)
%		(850,0.6332)
%		(875,0.5833)
%		(900,0.5419)
%		(925,0.5133)
%		(950,0.4912)%MIN
%		(975,0.5418)
%		(1000,0.5974)
%		};
%		\addlegendentry{$k=8$}
%		
%		\addplot[mark = oplus, smooth, blue, thick] plot coordinates {
%		(800,1.3722)
%		(825,1.1844)
%		(850,1.0268)
%		(875,0.9781)
%		(900,0.9329)
%		(925,0.9117)
%		(950,0.8922)
%		(975,0.8332) %MIN
%		(1000,0.8818)
%		};
%		\addlegendentry{$k=16$}
%		
%		\end{axis}
%		\end{tikzpicture}	
%	
%	\caption[2]{The best Compression Rate for overall data for $n_{o} = 1024$ for various variables of $n_{b}$}
%	\label{fig:result2}
%\end{figure}

\begin{table}[t]
	\caption{The Best Compression Rate for overall data, cloud and client for $n_{o} = 1024$ for different values of $k$.}
	\vspace{-1.3em}
	\begin{center}
		\begin{tabular}{|c|c|c||c|c|c|}
			\hline
			$n_{b}$ & $k$ & $\threshold$ & $\compratio$& $\ucr$ & $\ccr$ \\
			\hline
			900& 2 & 85 & 0.5527 & 0.2482 & 0.3158\\
			900& 4 & 44 & 0.5341 & 0.2119 & 0.3012\\
			900& 8 & 25 & 0.5419 & 0.1773 & 0.3645 \\
			900& 16 & 16 & 0.8493 & 0.1468 & 0.7025\\
			900& 32 & 6 & 0.9329 & 0.1159 & 0.08168 \\
			\hline
			950& 2 & 115 & 0.6583 & 0.4423 & 0.2159\\
			950& 4 & 59 & 0.5927 & 0.4038 & 0.1889 \\
			950& 8 & 31 & 0.4912 & 0.3645 & 0.1266 \\
			950& 16 & 15 & 0.8922 & 0.3218 & 0.6705\\
			950& 32 & 7 & 1.0657 & 0.2894 & 0.7352 \\
			\hline
		\end{tabular}
		\label{tab1}
		\vspace{-2.5em}
	\end{center}
\end{table}	

%	Table~\ref{tab1} includes a summary of the best compression rate for different values of $k$ and $n_{b}$. This table shows that by increasing the value of $k$ or $n_{b}$ the compression rate in the client is decreased, as the compression rate in the client is only dependent on the umber of eliminations. However, changing the value of $k$ and $n_{b}$ changes the compression rate of the cloud in a way that there is an optimal point. for example, for $n_{b} = 900$, the optimal point is at $k=4$, but the optimal point at $n_{b} = 950$ occurs at $k=8$.

% To allow compilation of the file
% !TeX root = ./../main.tex
\vspace{-5pt}
\section{Conclusion}\label{sec:conclusion}
We presented \name, a protocol that allows Cloud Storage Providers to carry out deduplication across data uploaded by multiple clients, while introducing a level of uncertainty which provides privacy for the data sent by the individual clients.
This injection of uncertainty is carried out by each client individually by transforming the data prior to upload and %maintaining descriptions of these modifications locally.
storing a compact description of such transformations locally.
%For \name, we relied on random deletion of symbols in a chunk of data, where each chunk has a different deletion pattern.
%For the Cloud provider, this is similar to experiencing a deletion channel, using terminology from information theory, which is known to be hard to recover data from and which has been used for practical physical layer security solutions. \daniel{need to add a few references here}.
%The CSP can then carry out deduplication of the data chunks sent by each individual client.
We propose an improvement on the concept of generalized deduplication to increase its compression potential, that consists of allowing the CSP to swap and change values of the records it receives from clients.
%In particular, we rely on swaps and substitution of symbols in the chunks to cluster data around a list of interesting data chunks.
Our numerical results show that \name reduces the amount of data stored in the Cloud, in the local device and even in the system as a whole, while providing a high degree of uncertainty regarding the data uploaded by each client.
%We also show that the data across the entire system experiences a net compression, i.e., the overall cost of storage is lowered.
%We expect that this results will be improved when considering a larger number of clients and stored data in the system, as the Cloud component will be able to deduplicate data more effectively.
A side advantage of \name is that it can protect from side channel attacks from malicious clients trying to gain knowledge about the data stored by other clients in the Cloud.
% This is due to the transformations introduced by each individual client.
Future work will consider malicious adversaries and clients that do not reveal the original size of the chunk and only comply with the expected basis size of the Cloud.
This added uncertainty requires further analysis as the attacker would not have the information about the chunk's original size as prior knowledge.

%beyond elimination operations and even consider that the original chunk size is a choice by each client with the goal to match a given basis size.
%This can introduce an added uncertainty as even the starting chunk size could be part of the hidden information at the client side.
%Analysing the trade-off between data stored in the Cloud and data stored in the client by including more transformation operations in the client will also be analyzed in future work.

%\elena{we should have a standard bibliography = separate bib file with items in the standard format!}

\ifpublish{
\section{Acknowledgement}
This work was partially financed by the SCALE-IoT project (Grant No. DFF - 7026-00042B) granted by the Danish Council for Independent Research, the Aarhus Universitets Forskningsfond Starting Grant Project AUFF-2017-FLS-7-1, Aarhus University's DIGIT Centre, the ERC Horizon 2020 Grant No 669255 (MPCPRO), the ELLIIT grant and the Swedish Foundation for Strategic Research, grant RIT17-00.
}
Finally, the authors would like to kindly thank Prof. Claudio Orlandi for useful discussions and insights on our work. 
 
 \fi
\bibliographystyle{IEEEtran}
\bibliography{main}

%\begin{thebibliography}{99}
%	\bibitem{rasmus}
%	Vestergaard, Rasmus, Daniel E. Lucani, and Qi Zhang. "Generalized Deduplication: Lossless Compression for Large Amounts of Small IoT Data." European Wireless 2019; 25th European Wireless Conference. VDE, 2019.
%	\bibitem{amir}
%	Amir, Amihood, Estrella Eisenberg, and Ely Porat. "Swap and mismatch edit distance." Algorithmica 45.1 (2006): 109-120.
%\end{thebibliography}

\end{document}